\documentclass[twocolumn,showpacs,amsmath,aps]{revtex4}
\usepackage{txfonts}
\usepackage{graphicx,amssymb,mathrsfs,bm,color,times}

\newcommand{\nl}{\nonumber \\}
\newcommand{\be}{\begin{equation}}
\newcommand{\ee}{\end{equation}}
\newcommand{\bea}{\begin{eqnarray}}
\newcommand{\eea}{\end{eqnarray}}
\newcommand{\bsube}{\begin{subequations}}
\newcommand{\esube}{\end{subequations}}

\newcommand{\Eq}[1]{Eq.\,(\ref{#1})}

\newcommand{\la}{\langle}
\newcommand{\ra}{\rangle}

\newcommand{\dg}{\dagger}
\makeatletter 
    
    \newcommand{\Rmnum}[1]{\expandafter\@slowromancap\romannumeral #1@}
\makeatother

\begin{document}
\title { Deterministic creation and stabilization of entanglement
         in circuit QED by homodyne-mediated feedback control }

\author{Zhuo Liu, L\"ulin Kuang, Kai Hu, Luting Xu,
        Suhua Wei, Lingzhen Guo \footnote{E-mail:guolingzhen@mail.bnu.edu.cn}
        and Xin-Qi Li \footnote{E-mail: lixinqi@bnu.edu.cn}}
\affiliation{Department of Physics, Beijing Normal University,
Beijing 100875, China}

\date{\today}

\begin{abstract}
In the solid-state circuit QED system and based on the homodyne
measurement in dispersive regime,
we demonstrate that a homodyne-current-based feedback
can create and stabilize highly entangled two-qubit states
in the presence of moderate noisy environment.
Particularly, we present an extended analysis for the current-based
Markovian feedback, which leads to
an improved filtered-current-based feedback scheme.
We show that this is essential for us to achieve the desirable
control effect in present system.
\end{abstract}


\maketitle


\section{Introduction}
The circuit quantum electrodynamics (QED) \cite{Bla04,Sch04,Mooij04},
a solid-state analog of the conventional quantum optics cavity QED,
is a promising solid-state quantum computing architecture.
This architecture couples superconducting electronic circuit
elements, which serve as the qubits,
to harmonic oscillator modes of a microwave resonator,
which serve as a ``quantum bus"
that mediates inter-qubit coupling and facilitates
quantum measurement for the qubit state.
At the early stage of experiments, strong coupling between
the qubits and the quantum bus in the circuit QED system
should make it excellent platform for quantum control study.

In particular, quantum measurement in this system can be carried out
by operating in the dispersive limit
(where the detuning between the resonator and the qubit is
much larger than their coupling strength). In this limit, the
interaction induces a qubit state dependent frequency shift on the
resonator. By measuring the resonator output voltage with a homodyne
measurement, information about the qubit state is obtained.
In our present work, based on this homodyne measurement together with
the flexibility/advantage that feedback can be applied to either
the qubit or the microwave resonator mode,
we will illustrate how highly entangled state can be created and stabilized
from an initially separable state, by an appropriate feedback control.
It has been well known that quantum entanglement is one of the key ingredients
in order to realize quantum computation and quantum information processing.
Very recently, in the circuit QED system, interesting ideas were proposed
to {\it probabilistically} create entangled states by means of the
homodyne measurement alone \cite{Gam08},
following the general idea that measurement can be used as a nondeterministic
means of preparing quantum states that are otherwise difficult to obtain. However,
besides the drawback of probabilistic nature, this {\it measurement alone} approach
has no ability to stabilize/protect the obtained entangled state.
Generally speaking, entanglement degradation through uncontrolled coupling
with the environment remains a major obstacle in practice
\cite{Yu0409,Car04,Roo04}, which would limit the lifetime of entangled states
and demand efficient schemes to protect them.
In this context we remind that, owing to the remarkable progress in theory
and particularly in experiments on the real-time monitoring
and manipulation of individual quantum systems \cite{WM10},
the quantum feedback control technique may emerge as a natural
possible route to develop strategies to prepare entangled states
and prevent their deterioration.

In Ref.\ \cite{SM02}, for the optical cavity QED system,
the steady-state of two qubits (two-level atoms)
interacting simultaneously with a driving laser was shown to be entangled,
achieving a concurrence of 0.11 without measurement or feedback.
A later work, involving the use of a homodyne-mediated {\it direct} feedback
modulation of the laser that drives the cavity mode \cite{WJ05},
extended this study and showed that one can increase the
steady-state entanglement, i.e., the amount of the maximum
steady-state concurrence can be increased from 0.11 to 0.31.
In principle, maximally entangled states could be achieved \cite{Mab04},
by the use of Bayesian, or state estimation, feedback \cite{Doh99}.
This improvement, however, comes at the price of increasing the
experimental complexity, due to the challenging need for a real-time
estimation of the quantum state.
More recently, rather than the homodyne-mediated, a jump-based direct feedback
is demonstrated to allow the robust production of highly entangled states
in the optical cavity system \cite{Car0708}.
For the solid-state circuit QED, however, the single photon counting at microwave
frequencies is currently not possible, despite a couple of proposals and efforts
out there for doing it \cite{Rom09,Joh10}

In this paper we base the feedback on the more available homodyne
measurement scheme \cite{Teu09}.
Particularly, in Ref.\ \cite{SarMil05},
detailed analysis for measurement scheme to reach the quantum limit
was presented, and a deterministic creation of entanglement by using
the measurement based feedback
was illustrated in the absence of environmental decoherence.
In present work we extend the study in Ref.\ \cite{SarMil05} by taking into
account the environmental influence. We will show that all the four
Bell states can be deterministically created and stabilized
in the presence of moderate noisy environment,
with high quality compared to the existing results mentioned above.

\section{Model and Formalism}


In Fig.\ 1 we illustrate the schematic setup of the
solid-state superconducting circuit-QED system,
together with the measurement and feedback control idea.
The central section of superconducting coplanar waveguide plays the role of
the cavity and the superconducting qubits play the roles of the atoms.
The superconducting qubits are coupled to
a one-dimensional transmission line (1DTL) cavity
which acts as a simple harmonic oscillator.
The qubits, the 1DTL cavity and their mutual coupling can be well described
by the Jaynes-Cummings Hamiltonian:
\begin{equation}\label{JCH}
H=\omega_r a^\dagger a +\mathcal{E}(a^\dagger + a)
+\sum_{j=1,2}\left[\frac{\Omega_{j}}{2} \sigma_j^z
+ g_j(\sigma_j^-a^\dagger+\sigma_j^+a)\right].
\end{equation}
Here, the operators $\sigma^-_j(\sigma^+_j)$ and $a(a^\dagger)$ are,
respectively, the lowering (raising) operators for the $j$th qubit
and the cavity photons.
$\omega_r$ is the frequency of the cavity photon,
and $\Omega_{j}$ and $g_j$ are the $j$th
qubit transition energy and coupling strength to the cavity photon.
For simplicity and for the purpose to be clear later,
in this work we assume that
$\Omega_{1}=\Omega_{2}=\Omega$, and $g_1=-g_2=g$.
This implies that we assume two identical qubits
located in the cavity at places with the maximum field amplitude
and with a half-wavelength separation, as schematically shown in Fig.\ 1.
The $\mathcal{E}$-term in \Eq{JCH} stands for a microwave driving
to the cavity that is employed here for the task of measurement.
More explicitly, $\mathcal{E}=\epsilon e^{-i\omega_m t}+{\rm c.c}$,
where the frequency can differ from the cavity photon frequency,
i.e., $\Delta_r\equiv\omega_r-\omega_m\neq 0$.
In concern with the qubit-cavity coupling, we focus on the dispersive regime
\cite{Bla04,Sch04,Mooij04}, which corresponds to an energy detuning,
$\Delta=\omega_r-\Omega$, much larger than $g$.
In this limit, the canonical transformation,
$H_{\rm eff}\simeq U^{\dg}HU$, where
$U=\exp[\sum_j\lambda_j(a\sigma_j^+ -a^{\dg}\sigma_j^+)]$
with $\lambda_j=g_j/\Delta$,
yields (in the rotating frame with the driving frequency $\omega_m$)
\bea\label{H_eff}
H_{\rm eff}&\simeq& \Delta_r a^{\dg}a + \epsilon(a+a^{\dg})
  + (\Omega+\chi)J_z/2   \nl
 & & + \chi a^{\dg}a J_z + \chi(\sigma^+_1\sigma^-_2+\sigma^-_1\sigma^+_2) .
\eea
Here, we defined $\chi=g^2/\Delta$ and $J_z = \sigma^z_1+\sigma^z_2$.

\begin{figure}
 \center
 \includegraphics[scale=0.5]{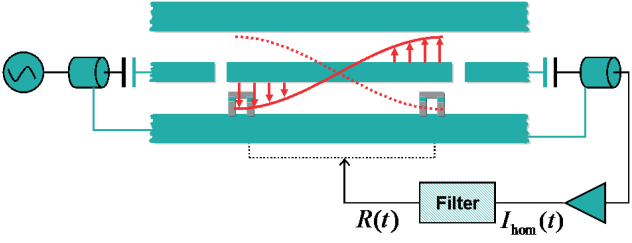}
 \caption{
Schematic diagram of circuit QED together with a microwave
transmission measurement and a measurement-current-based feedback loop.
The Cooper-pair box qubits are fabricated inside a superconducting
transmission-line resonator and are capacitively coupled to the voltage
standing wave.  }
\end{figure}

%
In the circuit-QED system, the measurement is implemented via
a homodyne-type detection of the transmitted microwave photons
as schematically shown in Fig.\ 1.
For the composite system of the qubits plus the 1DTL cavity, the leakage
of photons is described by a Lindblad term $\kappa {\cal D}[a]\rho$
in master equation, where $\kappa$ is the leakage rate and
the Lindblad superoperator acting on the
reduced density matrix $\rho$ is defined by
${\cal D}[a]\rho=a\rho a^{\dg}-\frac{1}{2}\{a^{\dg}a,\rho\}$.
However, conditioned on the output (homodyne) current, i.e.,
$I_{\rm hom}(t)=\kappa\la a+a^{\dg} \ra_c(t)+\sqrt{\kappa}\xi(t)$,
there will be an additional unravelling term in the conditional
master equation, $\mathcal{H}[a]\rho_c \xi(t)$.
Here, $\la (\cdots) \ra_c(t)\equiv {\rm Tr}[(\cdots)\rho_c(t)]$
with $\rho_c(t)$ the conditional density matrix,
and $\mathcal{H}[a]\rho_c \equiv a\rho_c+\rho_c a^{\dg}
-\mathrm{Tr}[(a+a^{\dg})\rho_c]\rho_c$.
And, the quantum-jump related stochastic nature is characterized
by $\xi(t)$, a Gaussian white noise with properties of $E[\xi(t)]=0$
and $E[\xi(t)\xi(t')]=\delta(t-t')$, where $E[\cdots]$  means an ensemble
average over realizations of the noise.
Furthermore, in this work, we are interested in the regime of strongly
damped cavity, which enables us to adiabatically eliminate the cavity
degree of freedom \cite{WMJW9394}.
Qualitatively, from an observation on the effective coupling
$\chi a^{\dg}a J_z$ in \Eq{H_eff}, we can expect:
the measurement-backaction induced dephasing term
$\sim \mathcal{D}[J_z]\rho_c$,
the unravelling term $\sim\mathcal{H}[J_z]\rho_c \xi(t)$,
and the homodyne current $ I_{\rm hom}(t)\sim \la J_z\ra_c(t)+\xi(t)$.
Indeed, following the standard procedures of adiabatic elimination
\cite{WMJW9394}, one obtains the quantum trajectory equation (QTE)
involving only the qubit degree of freedom,
in a rotating frame with respect to the qubit Hamiltonian which reads
\begin{eqnarray}\label{QTE}
&&\dot{\rho}_c = -i\chi|\alpha|^2[J_z,\rho_c]
    +\sum_{j=1,2} \gamma_{j}\mathcal{D}[\sigma_j^-]\rho_c
    +\sum_{j=1,2}\frac{\gamma_{\phi j}}{2}\mathcal{D}[\sigma_j^z]\rho_c    \nl
  && +\gamma_{p}\mathcal{D}[\sigma_{1}^{-}-\sigma_{2}^{-}]\rho_c
   +\frac{\Gamma_{d}}{2}\mathcal{D}[J_{z}]\rho_c
   +\frac{\sqrt{\Gamma_m}}{2}\mathcal{H}[J_z]\rho_c\xi(t).
\end{eqnarray}
In deriving this result, we have used the assumption $\Delta_r=0$
and $\lambda_1=-\lambda_2=\lambda$.
In \Eq{QTE},
$\gamma_j$ and $\gamma_{\phi j}$ are the relaxation and dephasing rates
caused by the surrounding environment, while $\gamma_p=\kappa\lambda^2$
is a rate for the collective decay due to the Purcell effect.
Explicitly, the measurement-backaction induced dephasing rate
$\Gamma_d=8|\alpha|^2 \chi^2/\kappa$,
with $\alpha=-2i\epsilon/\kappa$.
Finally, the information gain rate $\Gamma_m$ in \Eq{QTE}
is in general related to the backaction dephasing rate through
the quantum efficiency, $\eta=\Gamma_m/(2\Gamma_{d})$.

\section{Homodyne-Mediated Feedback}


\subsection{Feedback Equation in Markovian Limit}

Based on the QTE, one can infer the conditional state $\rho_c(t)$.
In principle, one can then perform a state-estimate based feedback,
by an appropriate design for the feedback Hamiltonian.
This is the so-called Bayesian feedback scheme.
Another, much simpler, scheme is the direct feedback which depends linearly
on the detection signal, e.g., the homodyne current in the present case.
Indeed, the former scheme can result in an improvement over
the direct feedback, yet it comes at the cost of an increasing
experimental complexity due to the challenging need for a real-time
estimation of the quantum state.
In what follows, we focus on the direct current-based feedback scheme.
In general, we denote the feedback Hamiltonian as
$H_{\rm fb}(t)= I_{\rm hom}(t-\tau)\hat{F}$,
where $\tau$ stands for a possible time delay of the feedback,
while the homodyne current can be expressed as
$I_{\rm hom}(t)=\sqrt{\Gamma_m}\la J_z \ra_c(t)+\xi(t)$
after the adiabatic elimination of the cavity degree of freedom.
Performing this feedback leads to state change according to
$[\dot{\rho}_c(t)]_{\rm fb}=I_{\rm hom}(t-\tau){\cal K}\rho_c(t)$ ,
where ${\cal K}\rho_c(t)\equiv -i[\hat{F},\rho_c(t)]$.
In the Markovian limit by assuming $\tau=0$,
an appropriate combination of this equation with the QTE (\ref{QTE}) yields
\begin{eqnarray}\label{fb_QTE}
  &\dot{\rho}_c=&-i\chi|\alpha|^2\left[J_z,\rho_c\right]
  +\frac{\Gamma_d}{2}\mathcal{D}[J_z] \rho_c    \nonumber\\
  &&  +\sum_{j=1,2} \gamma_{j}\mathcal{D}[\sigma_j^-]\rho_c
  +\sum_{j=1,2}\frac{\gamma_{\phi j}}{2}\mathcal{D}[\sigma_j^z]\rho_c    \nl
  &&+\gamma_{p}\mathcal{D}[\sigma_{1}^{-}-\sigma_{2}^{-}]\rho_c
+ \frac{\sqrt{\Gamma_m}}{2} \mathcal{K}(J_z\rho_c+\rho_c J_z)   \nl
&& + \frac{1}{2} \mathcal{K}^2\rho_c
   +\left( {\frac{\sqrt{\Gamma_{m}}}{2}}
     \mathcal{H}\left[J_z\right] + {\cal K}  \right)\rho_c\xi(t).
\end{eqnarray}
Originally, this type of equation was obtained
by Wiseman and Milburn \cite{WM93}, by a careful interpretation
to the feedback as an Ito- or Stratonovich-type stochastic action.
Interestingly, we mention here an alternate (and equivalent) derivation
for this result. Since even in the Markovian limit the feedback can be
applied only after the measurement outcome,
in terms of discretized time interval we thus have
$\rho_c(t+dt)= e^{d{\cal K}(t)}\tilde{\rho}_c(t+dt)$,
where $\tilde{\rho}_c(t+dt)$ is the state after inferring
from the measurement record,
and the differential superoperator reads $d{\cal K}(t)=dI_{\rm hom}(t){\cal K}$.
Here, $dI_{\rm hom}(t)=\sqrt{\Gamma_m}\la J_z \ra_c(t)dt + dW(t)$,
while $dW(t)=\xi(t) dt$ is the Wiener increment that has the statistical
properties of $E[dW(t)]=0$ and $E[dW(t)dW(s)]=\delta(t-s) dt$.
Expanding this instantly fast feedback to second order and keeping
the infinitesimal increment to $O(dt)$ (noting that $[dW(t)]^2=dt$),
one then obtains \Eq{fb_QTE}.

\subsection{Entanglement Creation and Stabilization:
            Preliminary Result}

From now on we specify our study by using feedback to create
two-qubit entanglement, first based on \Eq{fb_QTE},
then on an improved scheme.
In particular, we will focus on feedback protection against the spontaneous
emission of qubits, while assuming the dephasing rates $\gamma_{\phi j}$
negligibly small for the transmon-type qubit \cite{Gam08}.
For two qubits, there are four maximally entangled states,
i.e., the Bell states:
$|\Psi_{\pm}\ra=\frac{1}{\sqrt{2}}(|00\ra \pm |11\ra)$,
and $|\Phi_{\pm}\ra=\frac{1}{\sqrt{2}}(|01\ra \pm |10\ra)$.
In what follows we only detail our analysis for $|\Phi_+\ra$,
and remain the others as brief discussion in the final concluding section.
Initially, we assume a separable state for the two qubits,
$(|0\ra+|1\ra)_1\otimes(|0\ra+|1\ra)_2
=(|00\ra+|11\ra)+(|10\ra+|01\ra)$,
which can be prepared easily by separate single bit rotations.
Within the scheme after adiabatic elimination of the cavity degree
of freedom, we see that the current $\la J_z \ra_c$ is zero
for qubit state $|10\ra+|01\ra$, and nonzero otherwise.
Then, we expect that the current-based feedback would
force the state towards the entangled state $|10\ra+|01\ra$,
if we exert a $J_x$-type flipping, i.e., apply a feedback
$H_{\rm fb}(t)=u I_{\rm hom}(t)J_x$, where $u$ characterizes
the feedback strength over the current-based modulation.

\begin{figure}
 \center
 \includegraphics[scale=0.9]{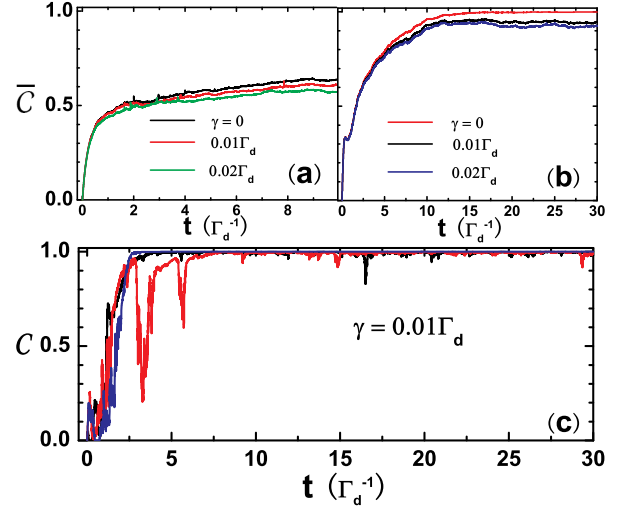}
 \caption{(a): Average concurrence over 500 trajectories based on \Eq{fb_QTE}
and a direct current feedback $H_{\rm fb}=u I_{\rm hom}(t)J_x$.
(b) and (c): Concurrence obtained by a feedback
$H_{\rm fb}=u \la J_z \ra_c(t)J_x$,
i.e., removing the noise in the homodyne current.
In (b) the average and in (c) the individual quantum trajectory concurrences
are illustrated, both showing considerable improvement over the result in (a).
Parameters: $\gamma_p=\Gamma_d$, $u=0.1$ in (a) and $u=1.0$ in (b) and (c).
Here, the choice of different feedback strength $u=0.1$ in (a) is
from a rough numerical optimization, since an increase of $u$ over this value
will give poorer result instead. }
\end{figure}

In Fig.\ 2(a), we illustrate the control result based on \Eq{fb_QTE},
where the average concurrence over 500 trajectories is shown.
Note that the concurrence, which is particularly useful
to characterize two-qubit entanglement in a mixed state\cite{Woo97},
is an effective measure in present context for the creation
and stabilization of $|\Phi_+\ra$,
which is in good consistence with the state fidelity,
$F(t)= {\rm Tr}[|\Phi_+\ra \la \Phi_+|\rho(t)]$.
Compared to Refs.\ \cite{SM02} and \cite{WJ05},
where the achieved concurrences are, respectively, 0.11 and 0.31,
we see that the result in Fig.\ 2(a) is good,
but not optimal for the following reasons.
First, we note that this improvement over the result in Ref.\ \cite{WJ05}
is largely owing to the different measurement scheme here,
which in the dispersive regime is in fact a continuous
quantum non-demolition (QND) measurement.
That is, here the measurement is a $J_z$-type,
while in Ref.\ \cite{WJ05} it is a $J_x$-type.
For current-based feedback, as heuristically discussed above,
the homodyne current $dI_{\rm hom}(t)$ contains,
not only the useful signal $\la J_z\ra_c(t)$, but also the harmful noise
(i.e. the $dW(t)$ term).
To this point, it is important to understand the distinct role of
the $dW(t)$ term in the homodyne current, when using it to update
the quantum state {\it versus} to perform the feedback.
In doing the former, it is informative;
while in doing the latter, it is useless and harmful.
In the current-based Markovian feedback equation, i.e., \Eq{fb_QTE},
the two Wiener increments were equated
and were mixed up in the final form \cite{WM93}.
We thus understand that it may lead to inefficient result sometimes.
As a quantitative support to this reasoning, in Fig. 2(b) and (c),
we show the result by adding a feedback,
$H_{\rm fb}=u \la J_z \ra_c(t)J_x$, straightforwardly into the
measurement record conditioned quantum trajectory equation.
We remind here that the noise in the homodyne current was removed.
In Fig.\ 2(b), the average concurrence by averaging over 500 quantum
trajectories is plotted, while in Fig.\ 2 (c) the concurrence of
individual quantum trajectories is illustrated.
Noticeably, essential improvement over the result in Fig.\ 2(a)
is achieved.

\subsection{Filtered-Current-Based Feedback: Improved Result}

As a matter of fact, the above ``improved" feedback,
$H_{\rm fb}=u \la J_z \ra_c(t)J_x$,
is a state-estimation feedback, since the $\la J_z \ra_c(t)$
is known {\it only} after knowing the state $\rho_c(t)$.
However, guided by it, we can refine the noisy homodyne current.
Following Ref.\ \cite{SarMil05}, we first low-pass filter
the measurement signal over a small time window,
$ R(t)=\frac{1}{\mathcal{N}}\int_{t-T}^te^{-\gamma_{\rm ft}(t-\tau)}
  dI_{\rm hom}(\tau)$,
where the factor $\mathcal{N}$ normalizes the maximum of the
smoothed signal $R(t)$ to unity.
Through this filtering procedure, we get in fact a crude but
efficient estimate of $\la J_z\ra_c(t)$.
Then, we condition the feedback with a power of the filtered measurement signal,
which would reduce the noise further in the estimate.
That is, we finally design a feedback Hamiltonian, $H_{fb}(t)= uR(t)^P J_x$,
adding directly into the measurement conditioned quantum trajectory equation,
where $P$ is the power to which the smoothed signal is raised.
In Fig.\ 3 we display the (a) average and (b) individual trajectory concurrences,
based on this filtered-current feedback control. We find that the result
is satisfactory, being comparable to that in Fig.\ 2(b) and (c).

\begin{figure}
 \center
 \includegraphics[scale=0.85]{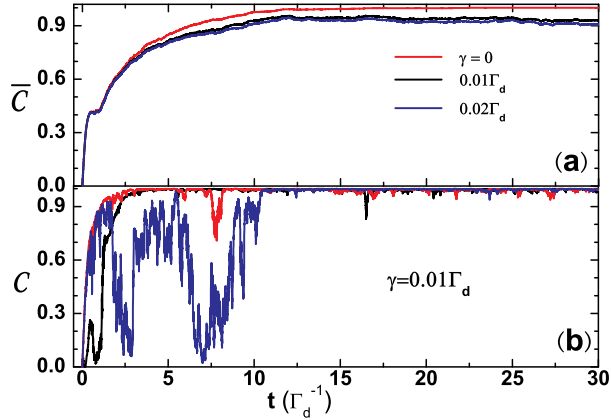}
 \caption{
(a) Average concurrence over 500 trajectories, and (b) the concurrence of
three representative individual trajectories, by applying a
filtered-current-based feedback control.
Parameters: $\gamma_p=\Gamma_d$, $u=10$,
$\gamma_{\rm ft}=0.006\Gamma_d$, and $T=2000\ast dt$.  }
\end{figure}

\begin{figure}
 \center
 \includegraphics[scale=0.8]{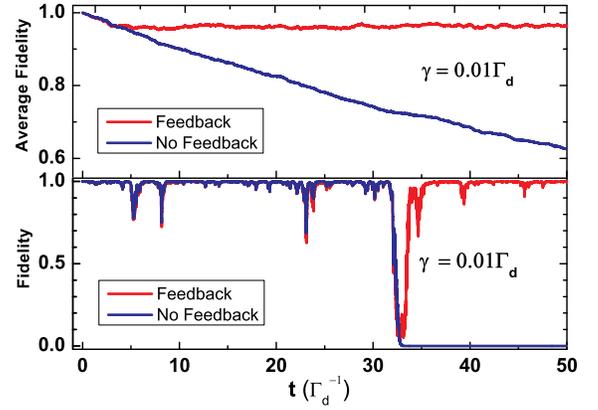}
 \caption{
Environment caused deterioration of the Bell state $|\Phi_+\ra$
in the absence of feedback protection.
In (a) the average fidelity (over 1000 trajectories)
and in (b) the fidelity of a representative
trajectory are presented (blue curves), while the result with
feedback is plotted here for comparison (red curves).
The feedback parameters are the same as in Fig.\ 3.   }
\end{figure}

\section{Entanglement Deterioration in the Absence of Feedback}

In addition to the {\it deterministic} creation of entanglement
achieved here, we remark that the other advantage of feedback is its ability
to {\it stabilize} the entanglement.
We illustrate this point more explicitly by the following.
Given the entangled state $|\Phi_+\ra$ having been achieved, for instance,
probabilistically by the {\it measurement alone} approach as proposed in
Ref.\ \cite{Gam08}, in Fig.\ 4 we plot the later fate of this state
in terms of its fidelity $F(t)={\rm Tr}[(|\Phi_+\ra \la \Phi_+|)\rho(t)]$,
in the absence of protection with feedback.
In Fig.\ 4(a) we show the average fidelity for the ensemble averaged state,
which is found to decay with time under the influence of environment.
By altering the coupling strength to environment, this decay may turn to
the remarkable phenomena of sudden death of entanglement \cite{Yu0409},
which were discussed under a broad class of dissipation models.
Meanwhile, in Fig.\ 4(b), the fidelity of representative individual trajectory
is shown. Interestingly, the sudden jump of fidelity to zero is observed,
which indicates a sudden disappearance of entanglement in our present case.
We stress that this is a real ``sudden death" of entanglement,
in the sense of an instantaneous disappearance of entanglement,
in contrast to the gradual death (i.e., with finite lifetime)
of the ensemble averaged state \cite{Yu0409}.
Anyhow, we may conclude, that even the entangled state (e.g. $|\Phi_+\ra$)
has been probabilistically created \cite{Gam08},
it will completely die, probably after some short surviving time,
but almost certainly with the increase of time,
in the absence of feedback protection.
Nevertheless, the feedback we discussed above can prevent this from happening.

\section{Discussion and Summary}

So far we have examined a feedback scheme of deterministic creation
and stabilization of entanglement. As a specific example, we considered
the state $|\Phi_{+}\ra=\frac{1}{\sqrt{2}}(|01\ra + |10\ra)$ in detail.
In practice,
the need of ``$J_x$ (=$\sigma^x_1+\sigma^x_2$)"-type feedback
can be realized by a modulation of the qubit Hamiltonian
parameters, as schematically indicated in Fig.\ 1,
or more simply, by modulating the microwave driving strength.
For the latter scheme, besides the measurement driving, one can apply an
additional control microwave driving, with amplitude of $\epsilon_c$
and frequency in resonance with the qubit transition energy.
It can be shown that this driving will induce an effective
$H_{\rm dr}=\lambda \epsilon_c J_x$ on the qubits,
if we properly set the qubits so that $g_1=g_2=g$, or equivalently,
$\lambda_1=\lambda_2=\lambda$.
Notice that in this case the Purcell term turns out to be
$\gamma_p {\cal D}[\sigma_1+\sigma_2]\rho$,
instead of $\gamma_p {\cal D}[\sigma_1-\sigma_2]\rho$ as in \Eq{QTE},
making the state $|\Phi_{+}\ra$ not invariant under its action.
However, one can make this term relatively small by appropriately
increasing the detuning between the cavity photon and the qubit.
Moreover, since this term only transits $|\Phi_{+}\ra$ to $|00\ra$,
yet $J_x|00\ra=|\Phi_{+}\ra$, we thus expect that the applied feedback
can well suppress its destructive effect.

Using similar method, one can deterministically create and stabilize
another entangled state $|\Phi_{-}\ra=\frac{1}{\sqrt{2}}(|01\ra - |10\ra)$.
This can be realized by a homodyne-mediated $\bar{J}_x$-type feedback,
where $\bar{J}_x\equiv\sigma^x_1-\sigma^x_2$.
This feedback can still be implemented by modulating either
the qubit Hamiltonian parameters or the microwave driving.
Finally, in concern with the other two Bell states,
$|\Psi_{\pm}\ra=\frac{1}{\sqrt{2}}(|00\ra \pm |11\ra)$,
we can create and stabilize them by a simple conversion scheme,
by noting that they are in fact related to
$|\Phi_{\pm}\ra=\frac{1}{\sqrt{2}}(|01\ra \pm |10\ra)$
by a single bit $\pi$-pulse $\sigma_x$ rotation.
Note that direct creation and stabilization of $|\Psi_{\pm}\ra$
is also possible, by alternatively setting $g_1=g_2=g$ and
an opposite detuning $\Omega_1-\omega_r=\omega_r-\Omega_2$.
Furthermore, an extension to protect arbitrary entangled states, say,
$|\tilde{\Phi}_{\pm}\ra=c|01\ra \pm d |10\ra$
and $|\tilde{\Psi}_{\pm}\ra=c|00\ra \pm d |11\ra$,
is straightforward, by simply replacing the feedback operator $J_x$
by an appropriate combination of $\sigma^x_1$ and $\sigma^x_2$.

Finally, we mention that the main problem with doing feedback
in circuit QED is the lack of efficient homodyne detection.
Currently, the way to perform homodyne and heterodyne detection
is to first amplify the signal before mixing it on a nonlinear
circuit element of some kind.
As a consequence, the extra noise added by the amplifier will
reduce the quantum efficiency and prohibit quantum limited feedback.
It seems that this situation is to be changed quickly, for instance,
by developing Josephson parametric amplifiers which can be
realized in superconducting circuits \cite{Teu09}.
In our present theoretical study, we did not include the
non-unit quantum efficiency into the homodyne detection of the field.
This treatment is largely owing to the following consideration.
After adiabatic elimination of the cavity photon degree of freedom,
the non-unit quantum efficiency of homodyne detection
will reduce the effective information-gain rate $\Gamma_m$ in \Eq{QTE}.
This implies an emergence of an extra non-unravelling dephasing term
in the quantum trajectory equation.
However, in the context of creation and stabilization of $|\Phi_+\ra$,
this term only results in dephasing among states
$|00\ra$, $|11\ra$, and $|01\ra + |10\ra$.
That is, it does not destroy the target state $|\Phi_+\ra$.
From the feedback principle discussed in Sec.\ III (B), we can also imagine
that this dephasing term only affects the detailed control dynamics,
but does not prevent the state towards the target state.
Therefore, for certain acceptable (not too low) quantum efficiency
of the homodyne detection, the present scheme of feedback should be
implementable in circuit QED.
This is seemingly a rare example that the control effect
does not sensitively depend on the measurement efficiency.
We have numerically examined the above reasoning,
for instance, by lowering the quantum efficiency to $\eta=0.8$,
and found very small change of the entanglement concurrence.

To summarize, in the solid-state circuit QED system and based on
the homodyne measurement in dispersive regime,
we analyzed the creation and stabilization of the highly
entangled Bell states by the use of a filtered-current-based feedback.
Compared to a few previous studies in similar cavity QED systems,
we (i) eliminate the drawback of probabilistic nature and the inability
of stabilization \cite{Gam08},
(ii) improve the control effect by enhancing the concurrence
to values higher than 0.9, in regard to Ref.\ \cite{WJ05} where the
improved concurrence of 0.31 over the former 0.11 was achieved,
and (iii) avoid the experimental difficulty in present system
for the jump-based feedback \cite{Car0708},
or the complexity for the state-estimation feedback \cite{Mab04}.


\vspace{1cm}
{\it Acknowledgements.}---
This work was supported by the National College Student Innovative
Experiment Program under No.\ 105102 \& 127012,
the NNSF of China under grants No.\ 101202101
\& 10874176, and the Major State Basic Research Project
under grant No. 2006CB921201.
XQL is grateful to Prof. G.J. Milburn for kind hospitality and
introducing to him the feedback control problem in circuit QED
during his visit to The University of Queensland,
and moreover, for informing the current status of measurement
in circuit QED which leads to the discussion of non-unit
measurement efficiency in the concluding section.
XQL also acknowledges the useful discussion with Prof. H.S. Goan
at the very beginning stage of this work.


\end{document}